\documentclass[superscriptaddress, 12pt, aps, prl,  preprint, longbibliography]{revtex4-1}
\usepackage{graphicx}
\usepackage{epstopdf}
\usepackage{color}
\usepackage{amsmath, amssymb}

\def\d{\mbox{d}}  
\def\e{\mbox{e}}  

\begin{document}

\title{Predicting Solvation Free Energies of Molecules and Ions via First-Principles and Machine-Learning Molecular Dynamics}
\author{Junting Yu}
\email{junting.yu@connect.ust.hk}
\affiliation{Department of Physics, Hong Kong University of Science and Technology, Hong Kong, China}
\author{Shuo-Hui Li}
\email{shuohuili@ust.hk }
\affiliation{Department of Physics, Hong Kong University of Science and Technology, Hong Kong, China}
\author{Ding Pan}
\email{dingpan@ust.hk}
\affiliation{Department of Physics, Hong Kong University of Science and Technology, Hong Kong, China}
\affiliation{Department of Chemistry, Hong Kong University of Science and Technology, Hong Kong, China}
\affiliation{IAS Center for AI for Scientific Discoveries, Hong Kong University of Science and Technology, Hong Kong, China}

\date{\today}

\begin{abstract}
The solvation free energy (SFE) of molecules and ions is a fundamental property governing their solvation behavior and solubility. Molecular simulations offer a route to compute SFEs using alchemical free energy methods, such as thermodynamic integration or free energy perturbation. However, 
these methods suffer from the infamous end-point singularity, which leads to numerical instability when atoms approach closely, a challenge that becomes particularly acute in ab initio and machine learning molecular dynamics simulations.
Here, we introduce the bubble method to calculate the SFEs of molecules and ions from first principles. Our approach avoids the end-state problem in both ab initio and machine learning molecular dynamics simulations and is applicable to molecules and ions of arbitrary shape. When calculating the SFEs of ions using periodic density functional theory, we incorporate corrections for the neutralizing background charge, spurious interactions between periodic images, and the vacuum-water interface potential. To validate our method, we successfully computed the SFEs of methane, methanol, water, and sodium ions using classical, ab initio, and machine learning molecular dynamics simulations.
Importantly, our method requires no experimental inputs or empirical data. This makes it particularly well-suited for studying systems under extreme conditions, such as high pressure-temperature environments or under nanoconfinement, situations where experimental investigations are challenging and classical force fields, typically parameterized under ambient conditions, may be unreliable.

\end{abstract}

\maketitle

\section{Introduction}
The maximum amount of a solute that can dissolve in a given solvent, defined as solubility, is a fundamental property driving processes in drug bioavailability, industrial extraction and purification, and geochemical systems. At the molecular level, this macroscopic phenomenon is governed by the solvation free energy (SFE), which quantifies the free energy change during the transfer of a solute from a vacuum into a solvent. In aqueous systems, this critical parameter is specifically termed the hydration free energy \cite{leung2009ab, luukkonen2020hydration, duarte2017approaches, shi2020absolute}.
Experimental measurements of SFEs are often challenging due to methodological uncertainties, particularly under elevated pressure (P) and temperature (T) conditions.
Thus, accurate molecular-level calculations offer a very valuable alternative for determining these properties.

Implicit solvation models are widely used to compute solvation free energy (SFE) due to their efficiency \cite{onufriev2008implicit, decherchi2015implicit, roux1999implicit}. SFE is split into polar and non-polar parts, typically modeled using the generalized Born (GB) approximation and solvent-accessible surface area, respectively \cite{bashford2000generalized}. 
These methods may also integrate with electronic structure calculations like density functional theory (DFT) \cite{mathew2014implicit, sinstein2017efficient}. However, they rely on empirical parameters and ignore molecular solvent structure, limiting their accuracy under non-ambient conditions such as extreme pressures, temperatures, or nanoconfinement.

With growing computational resources, explicit solvent models using molecular dynamics (MD) or Monte Carlo (MC) simulations are now widely used to compute solvation free energies (SFEs) via alchemical free energy methods, e.g.,  thermodynamic integration or free energy perturbation  \cite{lu2001accuracy, shivakumar2010prediction, mezei1987finite, khanna2020performing, wang2022automated}. Databases such as FreeSolv provide reference SFEs for neutral molecules from classical force fields and experiments \cite{duarte2017approaches}. A key challenge for those methods is the end-point singularity: at full decoupling,  Coulomb and van der Waals interactions diverge when atoms approach closely \cite{beutler1994avoiding}. This is addressed in classical force fields by soft-core potentials, which smooth interactions via an intermediate parameter \cite{beutler1994avoiding,steinbrecher2011soft}. These have been  implemented in popular codes like GROMACS and LAMMPS \cite{berendsen1995gromacs, thompson2022lammps}. 

In first-principles DFT or machine learning calculations, however, the implementation of analogous soft interactions remains an outstanding methodological challenge.
In DFT calculations, electronic energies often fail to converge in self-consistent field methods when atoms are too close. 
Machine learning potentials generally lack explicit physical forms for Coulomb and van der Waals interactions. Furthermore, their training data, derived from DFT or other high-level methods, rarely include configurations with extremely close atoms. As a result, machine learning potentials give unphysical energies when applied to such out-of-distribution geometries.

An alternative method to solve the end-state problem is to construct a fictitious cavity around the solute molecule, thereby decomposing the alchemical free energy calculation into a series of sub-processes \cite{li2017computational, li2018computational}. This cavity method has been  applied to compute SFEs of monoatomic ions using first-principles MD simulations \cite{Lin2024Ion}. However, for solute molecules or ions lacking spherical symmetry, this approach may exhibit numerical instability when employed in first-principles MD frameworks.

Here, we introduce the bubble method to calculate the SFEs of molecules and ions from first principles. Our method can avoid the end-point singularity in the alchemical free energy methods, and is applicable to a wide range of molecular and ionic shapes and can be directly implemented in both ab initio and machine learning MD simulations. For calculating the SFEs of ions within periodic DFT, we incorporate corrections for the neutralizing background charge, spurious interactions between periodic images, and the vacuum-water interface potential. Importantly, our method does not rely on any experimental inputs or empirical data. This makes it particularly well-suited for studying systems under extreme conditions, such as high pressure-temperature environments or under nanoconfinement, where experimental investigations are challenging and classical force field, typically parameterized under ambient conditions, may be unreliable.

\section{results and discussion}

\subsection{The bubble method}
 
Here we apply the thermodynamic integration (TI) method to calculate the SFE, which creates a fictitious Hamiltonian $H(\lambda)$ as a function of the coupling parameter $\lambda$. 
For the end state at $\lambda = 0$, the Hamiltonian $H(0) = H_m + H_w$ represents the system where the solute molecule and solvent are non-interacting, while at $\lambda = 1$, the Hamiltonian $H(1) = H_{mw}$ describes the fully solvated state, where the solute is fully dissolved and interacting with the solvent.

Here we use the subscripts $m$, $w$ and $mw$ to 
denote the solute molecule, the solvent (e.g., water), and the solution system, respectively. The parameter $\lambda$, which runs from 0 to 1, represents a thermodynamic coupling process in which the interactions between the solute molecule and the solvent are gradually switched on.
The free energy change for this process can be computed using the thermodynamic integration formula:
\begin{equation}
\Delta G = \int_0^1 \left\langle \frac{\partial E(\lambda)}{\partial\lambda} \right\rangle \d\lambda,
\end{equation}
where the angle brackets $\langle \rangle$ denote an ensemble average, typically sampled via molecular dynamics or Monte Carlo simulations.

A common and straightforward choice for the Hamiltonian is the linear coupling form:
\begin{equation}
H(\lambda) = (1-\lambda)(H_m + H_w) + \lambda H_{mw}.
\end{equation}
Substituting this into the integral yields:
\begin{equation}\label{eqn:deltaG}
\Delta G = \int_0^1 \langle E_{mw} - E_m - E_w \rangle_{\lambda} \d\lambda,
\end{equation}
where $E_x$ represents the energy corresponding to the Hamiltonian $H_x$.

However, this approach suffers from an end-state problem at $\lambda = 0$. In certain configurations, particularly when atoms of the solute and solvent closely approach or overlap, the energy difference $E_{mw} - E_m - E_w$ can diverge. This singularity makes the integral in Eq.~\eqref{eqn:deltaG} challenging to evaluate numerically.

To solve this issue, we introduce an artificial repulsive two-body bubble potential, defined as 
$U_b(r,\lambda_e)=u(r)e^{\lambda_e}$.
This potential acts on every atom pair between the solute molecule and the solvent. Its purpose is to create a ``bubble expanding'' effect that pushes solvent molecules away from the solute. The intermediate Hamiltonian for this process is then given by
\begin{equation}
H(\lambda_e) = H_m+H_w+\sum_{i \in m, j \in w} U_b(r_{ij},\lambda_e),
\end{equation}
where the parameter $\lambda_e$ varies from $-\infty$ to a finite value $\lambda_E$.
The free energy change of this expanding process is 
\begin{equation}
\Delta G_e = \int_{-\infty}^{\lambda_E} \langle \sum_{i \in m, j \in w} U_b(r_{ij},\lambda_e) \rangle_{\lambda_e}\d\lambda_e,
\end{equation}
where we have used the fact that the derivative of \(U_b\) with respect to \(\lambda_e\) is \(U_b\) itself. At this stage, there is no interaction between solute and solvent molecules.
In numerical calculations, we can begin with a highly negative value of $\lambda_e$, making the sum of $U_b$ negligibly small. Then, we select a series of equally spaced $\lambda_e$ values for separate simulations. This approach avoids the potential divergence of $H_{mw}$.

Once the bubbles reach an appropriate size, we proceed to the next "switching" step. In this stage, we gradually turn on the interaction between the solute and solvent while simultaneously turning off all bubble potentials.

We choose the simple linear form
\begin{equation}
H(\lambda_s) = (1-\lambda_s)(H_m+H_w+H_b) + \lambda_s H_{mw},
\end{equation}
where $H_b$ is the sum of fully expanded bubble potentials in the last step. 
The free energy change becomes
\begin{equation}
\Delta G_s = \int_0^1 \langle E_{mw}-E_m-E_w-E_b \rangle_{\lambda_s} \d\lambda_s.
\end{equation}
If the bubbles expand sufficiently, the end-state problem at $\lambda_s = 0$ is avoided.

Alternatively, the switching step can be split into two separate steps: first, turning on the solute–solvent interaction, and then turning off the bubble potentials. The total free energy change from these two steps should be consistent with $\Delta G_s$, as free energy is a state function and independent of the path taken. However, this approach may introduce additional statistical uncertainties and increase computational cost.

The SFE can be calculated by the sum of the two steps,
\begin{equation}
\Delta G = \Delta G_e + \Delta G_s.
\end{equation}
The remaining question lies on the form of $u(r)$, which, in principle, does not affect the final result of SFE.
An important requirement is that $u(r)$ must be finite when $r$ is small; 
otherwise the end-point singularity persists. 
In this study we mainly used the form 

$u(r)=\e^{-r/r_0}$, 
where $r_0=0.2$ \AA$^{-1}$. 
This is similar to the Buckingham potential, 
where the $r^{-6}$ term is set to 0.
We also tested other parameter values and potential forms, such as the Gaussian form $u(r) = \e^{-Br^2}$, the exponential cubic form $u(r)=\e^{-Br^3}$ and the Fermi-Dirac potential $u(r)=1/(\e^{B(r-r_0)}+1)$, in classical force field simulations. 

The details of these tests are shown in the supporting information. 
Our results show that the choice of parameters and potential forms does not significantly affect the final SFE.
In the previous cavity method, the spherical cavity centered at the molecule was required to enclose the entire molecule \cite{li2017computational, li2018computational}. This poses challenges for numerical stability and statistical sampling when the molecule is non‑spherical and the simulation box is small, as is often the case in AIMD simulations.

\subsection{Solvation free energies of Neutral molecules}

We selected three neutral molecules: methane, methanol, and water, to test our bubble method.
While methane does not dissolve in water at ambient conditions, the other two do .
Table \ref{tbl:SFE} shows the calculated results by the classical OPLS-AA force field \cite{jorgensen1996development}, and PBE-D3 \cite{perdew1997generalized} and revPBE-D3 \cite{zhang1998comment} exchange correlation functionals,  in comparison with experimental values at T = 300 K and P = 1 bar. 

\begin{table}[!ht]
\caption{Solvation free energies of neutral molecules and Na$^+$ obtained by the classical OPLS-AA force field, and PBE-D3 and revPBE-D3 exchange correlation functionals.  Experimental values of the neutral molecules are from Ref.  \cite{ben1984solvation}. The experimental value of Na$^+$ is inferred  from thermodynamic circles\cite{tissandier1998proton}. Simulation uncertainties were estimated by using the blocking error method \cite{flyvbjerg1989error} and error propagation rules (Unit: kcal/mol).
}
\centering
\begin{tabular}{c c c c c }
\hline
Molecule & OPLS-AA & PBE-D3 & revPBE-D3  & Expt. \\
\hline
CH$_4$ & 2.219(0.078) & 0.691(0.213) & 1.916(0.127) & 2.0 \\
CH$_4$ (sc) & 12.451(0.078) & 10.629(0.225) & 11.102(0.280) & - \\
CH$_3$OH & -4.287(0.152) & -10.510(0.309) & -6.334(0.275) & -5.1 \\
H$_2$O & -5.824(0.070) & -8.607(0.525) & -7.054(0.337)  & -6.3 \\
Na$^+$ & - & -91.21(1.06) & -88.10(0.85)& -101.29 \\
\hline
\end{tabular}
\label{tbl:SFE}
\end{table}

We found that the OPLS-AA force field yields the highest SFE, followed by revPBE-D3 and PBE-D3. Among these, OPLS-AA and revPBE-D3 show closer agreement with experimental values.

The non-polar methane molecule has a positive SFE, meaning that it is more energetically favorable for methane to remain in the dilute gas phase than to dissolve in water.
The polar water and methanol molecules containing an -OH group have negative SFEs, and are easier to be dissolved into water.
For the methane molecule, we also calculated its SFE at a supercritical condition (T=1000 K, P=1 GPa). 
Results of three methods are around 12 kcal/mol, which are 10 kcal/mol higher than at ambient condition. 
This indicates that it is less favorable for methane to move from the dilute gas phase into water at this high temperature than at ambient condition.
This does not mean, however, that methane is less soluble at this supercritical condition than at ambient condition.

\subsection{H$_2$O with the neural network potential}

Machine learning techniques are now widely applied to construct potential energy surfaces using training data from first-principles calculations \cite{behler2007generalized, wang2018deepmd}. 
This involves mapping atomic configurations into symmetry-invariant descriptors, which are then used to predict the corresponding DFT-level energies and forces.
This machine learning approach achieves an accuracy comparable to first-principles methods while offering a significant improvement in computational efficiency \cite{Wang2026Machine}.

For example, Schran et. al. trained a committee neural network potential (CNNP) model for water using the DFT data obtained from the revPBE0-D3 functional \cite{schran2020committee}, which is one of most accurate functional for water systems \cite{Kapil2022first}.

Here we demonstrate that our bubble method can be seamlessly integrated with a machine learning potential. 
As an example, we applied the pretrained Generation-1 CNNP model, which treats nuclei classically, to calculate the SFE of a water molecule. While the CNNP model accurately describes bulk water, it performs poorly for a single water molecule (see the supporting information). This is because the model was primarily trained and optimized on bulk water properties. Specifically, the average standard deviation across the 8 committee models for a single water molecule is as high as 5 kcal/mol, which could significantly impact the accuracy of the SFE calculation,
so we directly calculated $E_m$ using DFT with the revPBE0-D3 functional, and used the CNNP model to obtain $E_w$ and $E_{mw}$.

The expanding and switching free energies were calculated as 4.588 $\pm$ 0.159 kcal/mol and -9.705 $\pm$ 0.357 kcal/mol, respectively, resulting in the SFE of -5.118 $\pm$ 0.391 kcal/mol. 
This value is slightly less negative than the experimental one (-6.3 kcal/mol).

\subsection{Solvation free energy of charged ions}
The calculation of SFEs for ions presents greater challenges than for neutral molecules, both experimentally and computationally. By definition, the absolute SFE of an ion corresponds to the free energy change associated with transferring the ion from the gas phase into an infinitely dilute solution. However, neither reference state is experimentally accessible as an equilibrium system: isolated ions in the gas phase tend to associate with counterions, while ions in solution inevitably exist at finite concentrations near charged interfaces. Consequently, absolute ionic SFEs cannot be measured directly but must instead be inferred through combinations of equilibrium measurements, spectroscopic data, and thermodynamic cycles \cite{hunenberger2011single}.

In periodic DFT simulations, additional complications arise. The presence of a net charge necessitates corrections for both the neutralizing background charge and spurious interactions between periodic images. Moreover, because the electrostatic potential differs between vacuum and bulk water, a charged ion crossing the vacuum–water interface experiences a potential step, contributing to the overall solvation free energy \cite{kathmann2011understanding}. As a result, the experimental SFE of an ion is typically decomposed into two components: a bulk contribution arising from ion–water interactions in the interior (the intrinsic SFE), and a surface contribution originating from the vacuum–water interfacial potential.
A range of methodological approaches have been developed to compute ionic SFEs, spanning classical force fields to DFT-based methods, and from continuum solvation models to fully atomistic explicit solvent representations. These studies have incorporated various correction schemes to address the aforementioned challenges \cite{smith2005free, leung2009ab, kelly2006aqueous, tomanik2020solvation, duignan2017real, duignan2017electrostatic, shi2020absolute, Lin2024Ion}.

When calculating the SFE of a charged ion using our method, we cannot directly evaluate the energy difference as in Eq.~\ref{eqn:deltaG}, because the charged and neutral systems have different energy references in DFT calculations with periodic boundary conditions. This issue is analogous to the treatment of charged defect formation energies in solids~\cite{Freysoldt2009fully}.
To address this, we introduce two correction terms:
\begin{equation}
    E_{\text{corr}} = E_{\text{corr}}^b + E_{\text{corr}}^{ii},
    \label{eqn:E_corr}
\end{equation}
where $E_{\text{corr}}^b$ corrects for the fictitious interaction between the ionic charge and the neutralizing background charge, and $E_{\text{corr}}^{ii}$ accounts for the electrostatic interaction between the ion and its periodic images.

Periodic DFT calculations introduce a uniform background charge to neutralize the net charge in the unit cell, which leads to the fictitious interaction between the net charge and the background charge in the electrostatic energy (referred as the Hartree energy here).
The solvation Hartree energy difference in Eq.~\ref{eqn:deltaG} is calculated as: 
\begin{equation}
  \Delta  E^H = E^H_{iw} - E^H_i - E^H_w,
  \label{eqn:DeltaEH}
\end{equation}
where $E^H_{iw}$ is the total electrostatic energy of the ion in water ($iw$),  $E^H_i$ is the ion energy ($i$), and $E^H_w$ is for pure water ($w$).
Considering the background neutralizing charge, we can decompose those Hartree energies as
\begin{align}
E^H_{iw} &= E^H_{iw-iw} + 2E^H_{iw-b} + E^H_{b-b} \label{eqn:EH_iw}\\
E^H_i &= E^H_{i-i} + 2E^H_{i-b} + E^H_{b-b} \label{eqn:EH_i}\\
E^H_w &= E^H_{w-w} \label{eqn:EHw}
\end{align}
where the subscript $b$ refers to the background charge.
In periodic DFT, each Hartree term can be calculated as
the electrostatic energy between the charge density $\rho_x$ and potential $V_y$ generated by the charge density $\rho_y$,
\begin{equation}
E^H_{x-y} = \frac{1}{2}\int_{\Omega} \rho_a(\mathbf{r}) V_y(\mathbf{r}) \d\mathbf{r} = \frac{2\pi}{\Omega}\sum_{\mathbf{G}}\frac{\tilde{\rho}_x(\mathbf{G})\tilde{\rho}_y(\mathbf{G})}{G^2},
\label{eqn:x_y_interaction}
\end{equation}
where the integral is calculated in one unit cell with the volume of $\Omega$, and $\tilde{\rho}(\mathbf{G})$ is the Fourier transform of $\rho (\mathbf{r})$ in reciprocal space. Note that the charge density $\rho (\mathbf{r})$ contains both electronic and nuclear charges.
It is obvious that $E^H_{x-y}=E^H_{y-x}$.
It is worth noting that although $E^H_{iw}$ and $E^H_i$ are finite, each single term on the right hand side of Eqs. \ref{eqn:EH_iw} and \ref{eqn:EH_i} diverges due to the non-zero net charge ($\tilde{\rho}_x(\mathbf{G}=0) \neq 0 \; \text{for} \;  x \in \{iw, i, b\}$).

After substituting Eqs. \ref{eqn:EH_iw}, \ref{eqn:EH_i} and \ref{eqn:EHw} into Eq. \ref{eqn:DeltaEH}, we obtain
\begin{equation}
\Delta E^H 
= (E^H_{iw-iw}-E^H_{i-i}-E^H_{w-w}) + 2(E^H_{iw-b}-E^H_{i-b})
\label{eqn:DeltaEH-2}
\end{equation}
where the first term is the true electrostatic energy difference during solvation, and the second term accounts for the fictitious interactions with the background charge.
Thus, the correction term $E_{corr}^b$ in Eq. \ref{eqn:E_corr} is
\begin{gather}
E_{corr}^b = -2(E^H_{iw-b}-E^H_{i-b})  \nonumber \\ 
= -\int_{\Omega} \rho_b(\mathbf{r}) [V_{iw}(\mathbf{r}) -V_{i}(\mathbf{r}) ] \d\mathbf{r}= q(\overline{V}_{iw}-\overline{V}_i)
\end{gather}
where the background charge $\rho_b(\mathbf{r}) = -\frac{q}{\Omega}$, and $q$ is the net charge of the ion in water.
The potentials $\overline{V}_{iw}$ and $\overline{V}_i$
are the spatially averaged local potential in the ion-water and isolated ion systems, respectively, both of which are readily available in DFT calculations.

The other correction term $E_{\text{corr}}^{ii}$ comes from the electrostatic interaction between the ion and its periodic images. 
Experimental measurements are typically performed at low ion concentrations and subsequently extrapolated to the infinite dilution limit.
In periodic DFT calculations,
a low ion concentration requires a large unit cell, which is computationally expensive.
The first term in Eq. \ref{eqn:DeltaEH-2} is 
\begin{equation}
E^H_{iw-iw}-E^H_{i-i}-E^H_{w-w} = \frac{1}{2}\int_{\Omega} [\rho_{iw}(\mathbf{r})V_{iw}(\mathbf{r}) - \rho_{w}(\mathbf{r})V_{w}(\mathbf{r}) - \rho_{i}(\mathbf{r})V_{i}(\mathbf{r})] \d\mathbf{r}.
\label{eqn:true-DeltaEH}
\end{equation}
We consider the true electrostatic energy difference during solvation without ion-ion interactions as:
\begin{equation}
E^H_{int} = \frac{1}{2}\int_{\Omega} [\rho_{iw}(\mathbf{r})-\rho_w(\mathbf{r})]V_w(\mathbf{r})\d\mathbf{r}, 
\end{equation}
which represents the interaction energy between the charge density difference induced by solvation and the electrostatic potential of bulk water. Thus, the correction
\begin{equation}
E_{corr}^{ii} = E^H_{int} - (E^H_{iw-iw}-E^H_{i-i}-E^H_{w-w}) = \frac{1}{2}\int_{\Omega} [\rho_i(\mathbf{r})V_i(\mathbf{r}) - \rho_{iw}(\mathbf{r})\delta V(\mathbf{r})]\d\mathbf{r},
\end{equation}
where $\delta V(\mathbf{r}) = V_{iw}(\mathbf{r})-V_w(\mathbf{r})$.

The expanding, switching free energy values with the two corrections in Eq. \ref{eqn:E_corr} are listed in Table \ref{tbl:Na}.

Because there is only one ion in the unit cell, the bubble does not need to expand significantly, resulting in a small expansion free energy ($\Delta G_e$). Additionally, the background charge correction ($E_{\text{corr}}^b$) is also small, suggesting that the spatially averaged local potentials differ little between the isolated and dissolved ions.

The switching contribution, accounting for $\sim$160 kcal/mol to the SFE, is the dominating part. 
The ion-ion correction term, $E_{\text{corr}}^{ii}$, further reduces the SFE by about 20 kcal/mol. 

\begin{table}[!ht]
\caption{Contributions to the solvation free energy of Na$^+$. All values are in kcal/mol, except for the surface potential (V).}
\centering
\begin{tabular}{c c c c c c c }
\hline
Functional & $\Delta G_e$ & $\Delta G_s$ & $E_{corr}^b$ & $E_{corr}^{ii}$ & $V_s (V)$& $\Delta G_{solv}$ \\
\hline
PBE+D3 & 1.22(0.09) & -162.35(0.53) & 0.35(0.07) & -21.82(0.98) & 3.963 & -91.21(1.06) \\
revPBE+D3 & 0.82(0.05) & -160.667(0.40) & -0.11(0.01) & -14.88(0.91) & 3.761 & -88.10(0.85) \\ 
\hline
\end{tabular}
\label{tbl:Na}
\end{table}

\subsection{Surface potential of vacuum-water interface}

In the experimental measurement of the ion SFE, the solvation process involves crossing the vacuum-water interface, and there is a potential jump at the water surface. This is also referred to as the mean inner potential of water in certain contexts \cite{harscher1998inelastic, sellner2014matter, yesibolati2020mean}.
Thus, the experimental ion SFE contains the bulk and the surface contributions:
\begin{equation}
\Delta G_{ion} = \Delta G_{bulk} + \Delta G_{surf}.
\end{equation}
The surface contribution ($\Delta G_{surf}$) is proportional to the potential jump ($V_{s}$) at the water surface, $\Delta G_{surf} = qV_{s}$. 
To calculate $V_{s}$ from MD simulations,
we built a water slab perpendicular to the $z$ direction.
We define an average electrostatic potential $\phi(z)$ along $z$-axis as
\begin{equation}
\phi(z) = \frac{1}{A_{xy}}\langle\iint V(x,y,z) \d x \d y \rangle,
\end{equation}
where $A_{xy}$ is the area of the $xy$-plane, the double integral represents the spatial average over the $xy$-plane,  $\langle \cdot \rangle$ denotes the ensemble average.
The electrostatic potential in the real space, $V(x,y,z)$ can be obtained by the total charge density using Poisson's equation or Green's function
\begin{gather}
\nabla^2V(\mathbf{r}) = -4\pi\rho_t(\mathbf{r}) \\
V(\mathbf{r}) = \int \frac{\rho(\mathbf{r}')}{|\mathbf{r-r'}|} \d V'
\end{gather}
We modifed the CP2K source code to calculate the spatial average $\iint \d x \d y / A_{xy}$ on the fly during MD simulations
\footnote{In the CP2K code, the electrostatic potential $V(x,y,z)$ is stored on a 3D uniform real-space grid, with columns along the $z$-direction distributed across different MPI processes. While $V(x,y,z)$ could in principle be written to a single file (e.g., a Gaussian cube file) for post-processing, this approach would require prohibitively large disk space. In our implementation, each process computes partial sums for its local $z$-columns, which are then sent to the master process for final summation and analysis.}.

\begin{figure}
\centering
\includegraphics[width=.8\linewidth]{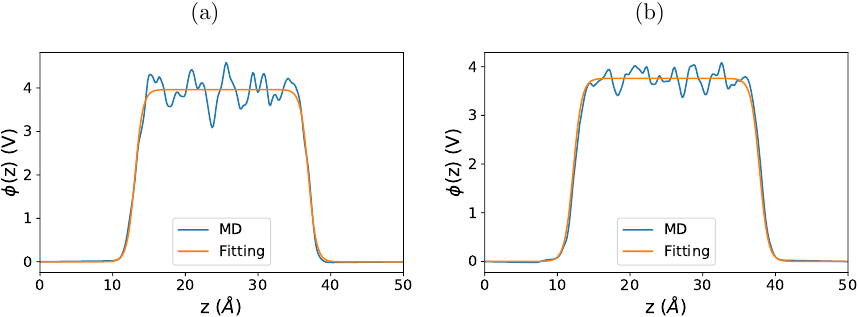}
\caption{
The $z$-dependence of the average electrostatic potential $\phi(z)$ for water slabs, derived from AIMD simulations and subsequent model fitting. The system consists of 128 water molecules in a unit cell with dimensions $12.41  \text{Å} \times 12.41  \text{Å} \times 50  \text{Å}$. The water molecules aggregate in the center of the cell, forming two distinct vacuum-water interfaces.
Two exchange-correlation functionals are compared: (a) PBE-D3, and (b) revPBE-D3.}
\label{fig:Vz}
\end{figure}

The $\phi(z)$ of the water slab obtained using PBE-D3 and revPBE-D3 are shown in Figure \ref{fig:Vz}. 
Given the symmetry of the curves about $z = 25$\,\AA, all data points are mapped to the left side for least squares fitting.
We used a logistic sigmoid (or hyperbolic tangent) function, as employed by Kathmann \cite{kathmann2011understanding}, to perform the fitting.
\begin{equation}
\hat{\phi}(z) = \frac{V_s}{1+\e^{-\beta(z-z_0)}} + V_0 =V_s \mathrm{Sigm}[\beta(z-z_0)] + V_0 = \frac{V_s}{2}(1+\mathrm{tanh}[\frac{\beta(z-z_0)}{2}]) + V_0,
\end{equation}
where 
$V_0$ is the reference potential, $z_0$ denotes the position of the symmetry axis for $\phi(z)$, and $\beta$ controls the steepness of $\phi(z)$ variation at the water surface.

The fitted surface potentials $V_s = 3.963\,\mathrm{V}$ (PBE-D3) and $3.761\,\mathrm{V}$ (revPBE-D3) align with previous computational and experimental values: $3.8\,\mathrm{V}$ (PBE)~\cite{sellner2014matter}, $3.63\,\mathrm{V}$ (PBE)~\cite{leung2010surface}, $3.5\,\mathrm{V}$ (indirect calculation)~\cite{cheng2009redox}, and $3.5\,\mathrm{V}$ (electron holography)~\cite{harscher1998inelastic,prozorov2017off}.

After considering the potential jump at the water surface, we obtained the SFE of $-91.21$ kcal/mol for Na$^+$ with the PBE-D3 functional and $-88.10$ kcal/mol with revPBE-D3. These values
are approaching the experimental value inferred from thermodynamic circles as shown in Table \ref{tbl:SFE}.

\section{conclusions}

We introduce a numerically stable method for computing SFEs of neutral molecules and ions that can be readily integrated into both ab initio and machine learning MD simulations. This method avoids the end-point singularity in thermodynamic integration and works for any molecular or ionic shape. For ions treated within periodic density functional theory, we explicitly account for three key electrostatic corrections: the neutralizing background charge, finite-size artifacts arising from periodic image interactions, and the vacuum–liquid interface potential. We validated our method by calculating the SFEs of methane, methanol, water, and Na$^+$ using classical, ab initio, and machine learning MD simulations.
The accuracy of the resulting solvation free energies is determined only by the quality of the underlying potential energy surface (force field parameters or exchange–correlation functional). Owing to its robustness and minimal reliance on system-specific empirical parameters, this method is especially advantageous for studying solvation under extreme thermodynamic conditions (e.g., high pressure and temperature) or in highly confined environments (e.g., nanopores, interfacial systems), situations where experimental data are scarce and conventionally parameterized classical force fields often become unreliable.

\section{Methods}

\subsection{Classical molecular dynamics} 
We performed classical molecular dynamics simulations using the LAMMPS package \cite{thompson2022lammps}. 
We used the OPLS-AA force field \cite{jorgensen1996development} and the SPC water model \cite{kusalik1994spatial}.  
We used the shake algorithm  to constrain their the bonds and angles of water molecules \cite{ryckaert1977numerical}. 
Our unit cell contains one solute and 512 water molecules with periodic boundary conditions. 
In MD simulations we equilibrated each system in isobaric-isothermal (NPT) ensemble for 100 ps, followed by another $\sim$100 ps NPT production run.
The timestep is 2 fs, the temperature was controlled at 298 K using the Nose-Hoover thermostat \cite{evans1985nose, Hoover1985Canonical}, and pressure was set at 1 bar with the Nose-Hoover barostat \cite{tuckerman2006liouville}. 

\subsubsection{Ab initio molecular dynamics} 
We performed ab intio molecular dynamics (AIMD) simulations using the CP2K Quickstep package \cite{kuhne2020cp2k,vandevondele2005quickstep}. 
We used the Goedecker–Teter–Hutter (GTH) norm-conserving pseudopotentials to describe core electrons \cite{goedecker1996separable,hartwigsen1998relativistic} and a triple-$\zeta$ doubly polarized (TZV2P) basis sets to expand the Kohn-Sham orbitals of valence electrons.
The electron density cutoff is 340 Ry. 
We used deuterium instead of hydrogen to make use of a larger time step of 0.5 fs. 
There are one solute and 64 water molecules in a cubic unit cell with periodic boundary conditions.
We applied the PBE\cite{perdew1997generalized} and revPBE\cite{zhang1998comment} exchange-correlation functionals with the D3 van der Waals (vdW) correction \cite{grimme2010consistent}.

We began each AIMD trajectory with an initial NPT simulation of 10 ps.
Temperature was regulated using the Bussi-Donadio-Parrinello thermostat ($\tau$ = 100 fs)~\cite{bussi2007canonical}, and pressure was controlled via modified Hoover's equations~\cite{martyna1994constant}. 
We calculated the average cell lengths from the final 5 ps of this trajectory, and then performed an NVT simulation for 20 ps, using the resulting unit cell dimensions and the same thermostat, to compute ensemble averages.

\subsection{Committee neural network potential} 
For the water SFE calculations with the committee neural network potentia (CNNP) model, we conducted the revPBE0-D3 calculations on both isolated molecules and bulk water (sampled every 100 MD steps) to compare energy differences between the hybrid functional and neural network potential, as shown in Table S5 in the supporting information. The bulk water has 64 molecules in the supercell with periodic boudary conditions. Our simulation parameters are the same as in the original CNNP paper\cite{schran2020committee}: 400 Ry plane-wave cutoff, cpFIT3 auxiliary basis set for ADMM~\cite{guidon2010auxiliary}, and 6 \AA\ Coulomb truncation for the Hartree-Fock exchange.

\section*{Acknowledgements}
This work was supported by the Hong Kong Research Grants Council (RGC) (Projects GRF-16301723, GRF-16306621, GRF-16302423, and GRF-16310225), and National Natural Science Foundation of China/RGC Joint Research Scheme (N\_HKUST664/24).
Part of this work was carried out using computational resources from the National Supercomputer Center in Guangzhou, China, and the X-GPU cluster supported by the RGC Collaborative Research Fund C6021-19EF.

\bibliography{ref}

\begin{thebibliography}{10}
\expandafter\ifx\csname url\endcsname\relax
  \def\url#1{\texttt{#1}}\fi
\expandafter\ifx\csname urlprefix\endcsname\relax\def\urlprefix{URL }\fi
\providecommand{\bibinfo}[2]{#2}
\providecommand{\eprint}[2][]{\url{#2}}

\bibitem{leung2009ab}
\bibinfo{author}{Leung, K.}, \bibinfo{author}{Rempe, S.~B.} \&
  \bibinfo{author}{von Lilienfeld, O.~A.}
\newblock \bibinfo{title}{Ab initio molecular dynamics calculations of ion
  hydration free energies}.
\newblock \emph{\bibinfo{journal}{The Journal of chemical physics}}
  \textbf{\bibinfo{volume}{130}}, \bibinfo{pages}{204507}
  (\bibinfo{year}{2009}).

\bibitem{luukkonen2020hydration}
\bibinfo{author}{Luukkonen, S.}, \bibinfo{author}{Levesque, M.},
  \bibinfo{author}{Belloni, L.} \& \bibinfo{author}{Borgis, D.}
\newblock \bibinfo{title}{Hydration free energies and solvation structures with
  molecular density functional theory in the hypernetted chain approximation}.
\newblock \emph{\bibinfo{journal}{The Journal of Chemical Physics}}
  \textbf{\bibinfo{volume}{152}}, \bibinfo{pages}{064110}
  (\bibinfo{year}{2020}).

\bibitem{duarte2017approaches}
\bibinfo{author}{Duarte Ramos~Matos, G.} \emph{et~al.}
\newblock \bibinfo{title}{Approaches for calculating solvation free energies
  and enthalpies demonstrated with an update of the freesolv database}.
\newblock \emph{\bibinfo{journal}{Journal of Chemical \& Engineering Data}}
  \textbf{\bibinfo{volume}{62}}, \bibinfo{pages}{1559--1569}
  (\bibinfo{year}{2017}).

\bibitem{shi2020absolute}
\bibinfo{author}{Shi, Y.} \& \bibinfo{author}{Beck, T.~L.}
\newblock \bibinfo{title}{Absolute ion hydration free energy scale and the
  surface potential of water via quantum simulation}.
\newblock \emph{\bibinfo{journal}{Proceedings of the National Academy of
  Sciences}} \textbf{\bibinfo{volume}{117}}, \bibinfo{pages}{30151--30158}
  (\bibinfo{year}{2020}).

\bibitem{onufriev2008implicit}
\bibinfo{author}{Onufriev, A.}
\newblock \bibinfo{title}{Implicit solvent models in molecular dynamics
  simulations: A brief overview}.
\newblock \emph{\bibinfo{journal}{Annual Reports in Computational Chemistry}}
  \textbf{\bibinfo{volume}{4}}, \bibinfo{pages}{125--137}
  (\bibinfo{year}{2008}).

\bibitem{decherchi2015implicit}
\bibinfo{author}{Decherchi, S.}, \bibinfo{author}{Masetti, M.},
  \bibinfo{author}{Vyalov, I.} \& \bibinfo{author}{Rocchia, W.}
\newblock \bibinfo{title}{Implicit solvent methods for free energy estimation}.
\newblock \emph{\bibinfo{journal}{European journal of medicinal chemistry}}
  \textbf{\bibinfo{volume}{91}}, \bibinfo{pages}{27--42}
  (\bibinfo{year}{2015}).

\bibitem{roux1999implicit}
\bibinfo{author}{Roux, B.} \& \bibinfo{author}{Simonson, T.}
\newblock \bibinfo{title}{Implicit solvent models}.
\newblock \emph{\bibinfo{journal}{Biophysical chemistry}}
  \textbf{\bibinfo{volume}{78}}, \bibinfo{pages}{1--20} (\bibinfo{year}{1999}).

\bibitem{bashford2000generalized}
\bibinfo{author}{Bashford, D.} \& \bibinfo{author}{Case, D.~A.}
\newblock \bibinfo{title}{Generalized born models of macromolecular solvation
  effects}.
\newblock \emph{\bibinfo{journal}{Annual review of physical chemistry}}
  \textbf{\bibinfo{volume}{51}}, \bibinfo{pages}{129--152}
  (\bibinfo{year}{2000}).

\bibitem{mathew2014implicit}
\bibinfo{author}{Mathew, K.}, \bibinfo{author}{Sundararaman, R.},
  \bibinfo{author}{Letchworth-Weaver, K.}, \bibinfo{author}{Arias, T.} \&
  \bibinfo{author}{Hennig, R.~G.}
\newblock \bibinfo{title}{Implicit solvation model for density-functional study
  of nanocrystal surfaces and reaction pathways}.
\newblock \emph{\bibinfo{journal}{The Journal of chemical physics}}
  \textbf{\bibinfo{volume}{140}}, \bibinfo{pages}{084106}
  (\bibinfo{year}{2014}).

\bibitem{sinstein2017efficient}
\bibinfo{author}{Sinstein, M.} \emph{et~al.}
\newblock \bibinfo{title}{Efficient implicit solvation method for full
  potential dft}.
\newblock \emph{\bibinfo{journal}{Journal of Chemical Theory and Computation}}
  \textbf{\bibinfo{volume}{13}}, \bibinfo{pages}{5582--5603}
  (\bibinfo{year}{2017}).

\bibitem{lu2001accuracy}
\bibinfo{author}{Lu, N.} \& \bibinfo{author}{Kofke, D.~A.}
\newblock \bibinfo{title}{Accuracy of free-energy perturbation calculations in
  molecular simulation. i. modeling}.
\newblock \emph{\bibinfo{journal}{The Journal of Chemical Physics}}
  \textbf{\bibinfo{volume}{114}}, \bibinfo{pages}{7303--7311}
  (\bibinfo{year}{2001}).

\bibitem{shivakumar2010prediction}
\bibinfo{author}{Shivakumar, D.} \emph{et~al.}
\newblock \bibinfo{title}{Prediction of absolute solvation free energies using
  molecular dynamics free energy perturbation and the opls force field}.
\newblock \emph{\bibinfo{journal}{Journal of chemical theory and computation}}
  \textbf{\bibinfo{volume}{6}}, \bibinfo{pages}{1509--1519}
  (\bibinfo{year}{2010}).

\bibitem{mezei1987finite}
\bibinfo{author}{Mezei, M.}
\newblock \bibinfo{title}{The finite difference thermodynamic integration,
  tested on calculating the hydration free energy difference between acetone
  and dimethylamine in water}.
\newblock \emph{\bibinfo{journal}{The Journal of chemical physics}}
  \textbf{\bibinfo{volume}{86}}, \bibinfo{pages}{7084--7088}
  (\bibinfo{year}{1987}).

\bibitem{khanna2020performing}
\bibinfo{author}{Khanna, V.}, \bibinfo{author}{Monroe, J.~I.},
  \bibinfo{author}{Doherty, M.~F.} \& \bibinfo{author}{Peters, B.}
\newblock \bibinfo{title}{Performing solvation free energy calculations in
  lammps using the decoupling approach}.
\newblock \emph{\bibinfo{journal}{Journal of Computer-Aided Molecular Design}}
  \textbf{\bibinfo{volume}{34}}, \bibinfo{pages}{641--646}
  (\bibinfo{year}{2020}).

\bibitem{wang2022automated}
\bibinfo{author}{Wang, F.} \& \bibinfo{author}{Cheng, J.}
\newblock \bibinfo{title}{Automated workflow for computation of redox
  potentials, acidity constants, and solvation free energies accelerated by
  machine learning}.
\newblock \emph{\bibinfo{journal}{The Journal of Chemical Physics}}
  \textbf{\bibinfo{volume}{157}}, \bibinfo{pages}{024103}
  (\bibinfo{year}{2022}).

\bibitem{beutler1994avoiding}
\bibinfo{author}{Beutler, T.~C.}, \bibinfo{author}{Mark, A.~E.},
  \bibinfo{author}{van Schaik, R.~C.}, \bibinfo{author}{Gerber, P.~R.} \&
  \bibinfo{author}{Van~Gunsteren, W.~F.}
\newblock \bibinfo{title}{Avoiding singularities and numerical instabilities in
  free energy calculations based on molecular simulations}.
\newblock \emph{\bibinfo{journal}{Chemical physics letters}}
  \textbf{\bibinfo{volume}{222}}, \bibinfo{pages}{529--539}
  (\bibinfo{year}{1994}).

\bibitem{steinbrecher2011soft}
\bibinfo{author}{Steinbrecher, T.}, \bibinfo{author}{Joung, I.} \&
  \bibinfo{author}{Case, D.~A.}
\newblock \bibinfo{title}{Soft-core potentials in thermodynamic integration:
  Comparing one-and two-step transformations}.
\newblock \emph{\bibinfo{journal}{Journal of computational chemistry}}
  \textbf{\bibinfo{volume}{32}}, \bibinfo{pages}{3253--3263}
  (\bibinfo{year}{2011}).

\bibitem{berendsen1995gromacs}
\bibinfo{author}{Berendsen, H.~J.}, \bibinfo{author}{van~der Spoel, D.} \&
  \bibinfo{author}{van Drunen, R.}
\newblock \bibinfo{title}{Gromacs: A message-passing parallel molecular
  dynamics implementation}.
\newblock \emph{\bibinfo{journal}{Computer physics communications}}
  \textbf{\bibinfo{volume}{91}}, \bibinfo{pages}{43--56}
  (\bibinfo{year}{1995}).

\bibitem{thompson2022lammps}
\bibinfo{author}{Thompson, A.~P.} \emph{et~al.}
\newblock \bibinfo{title}{Lammps-a flexible simulation tool for particle-based
  materials modeling at the atomic, meso, and continuum scales}.
\newblock \emph{\bibinfo{journal}{Computer Physics Communications}}
  \textbf{\bibinfo{volume}{271}}, \bibinfo{pages}{108171}
  (\bibinfo{year}{2022}).

\bibitem{li2017computational}
\bibinfo{author}{Li, L.}, \bibinfo{author}{Totton, T.} \&
  \bibinfo{author}{Frenkel, D.}
\newblock \bibinfo{title}{Computational methodology for solubility prediction:
  Application to the sparingly soluble solutes}.
\newblock \emph{\bibinfo{journal}{The Journal of chemical physics}}
  \textbf{\bibinfo{volume}{146}}, \bibinfo{pages}{214110}
  (\bibinfo{year}{2017}).

\bibitem{li2018computational}
\bibinfo{author}{Li, L.}, \bibinfo{author}{Totton, T.} \&
  \bibinfo{author}{Frenkel, D.}
\newblock \bibinfo{title}{Computational methodology for solubility prediction:
  Application to sparingly soluble organic/inorganic materials}.
\newblock \emph{\bibinfo{journal}{The Journal of chemical physics}}
  \textbf{\bibinfo{volume}{149}}, \bibinfo{pages}{054102}
  (\bibinfo{year}{2018}).

\bibitem{Lin2024Ion}
\bibinfo{author}{Lin, C.}, \bibinfo{author}{He, X.}, \bibinfo{author}{Xi, C.},
  \bibinfo{author}{Zhang, Q.} \& \bibinfo{author}{Wang, L.-W.}
\newblock \bibinfo{title}{Ion solvation free energy calculations based on
  first-principles molecular dynamics thermodynamic integration}.
\newblock \emph{\bibinfo{journal}{The Journal of Chemical Physics}}
  \textbf{\bibinfo{volume}{160}}, \bibinfo{pages}{184115}
  (\bibinfo{year}{2024}).

\bibitem{jorgensen1996development}
\bibinfo{author}{Jorgensen, W.~L.}, \bibinfo{author}{Maxwell, D.~S.} \&
  \bibinfo{author}{Tirado-Rives, J.}
\newblock \bibinfo{title}{Development and testing of the opls all-atom force
  field on conformational energetics and properties of organic liquids}.
\newblock \emph{\bibinfo{journal}{Journal of the American Chemical Society}}
  \textbf{\bibinfo{volume}{118}}, \bibinfo{pages}{11225--11236}
  (\bibinfo{year}{1996}).

\bibitem{perdew1997generalized}
\bibinfo{author}{Perdew, J.~P.}, \bibinfo{author}{Burke, K.} \&
  \bibinfo{author}{Ernzerhof, M.}
\newblock \bibinfo{title}{Generalized gradient approximation made simple (vol
  77, pg 3865, 1996)} (\bibinfo{year}{1997}).

\bibitem{zhang1998comment}
\bibinfo{author}{Zhang, Y.} \& \bibinfo{author}{Yang, W.}
\newblock \bibinfo{title}{Comment on “generalized gradient approximation made
  simple”}.
\newblock \emph{\bibinfo{journal}{Physical Review Letters}}
  \textbf{\bibinfo{volume}{80}}, \bibinfo{pages}{890} (\bibinfo{year}{1998}).

\bibitem{ben1984solvation}
\bibinfo{author}{Ben-Naim, A.} \& \bibinfo{author}{Marcus, Y.}
\newblock \bibinfo{title}{Solvation thermodynamics of nonionic solutes}.
\newblock \emph{\bibinfo{journal}{The Journal of chemical physics}}
  \textbf{\bibinfo{volume}{81}}, \bibinfo{pages}{2016--2027}
  (\bibinfo{year}{1984}).

\bibitem{tissandier1998proton}
\bibinfo{author}{Tissandier, M.~D.} \emph{et~al.}
\newblock \bibinfo{title}{The proton's absolute aqueous enthalpy and gibbs free
  energy of solvation from cluster-ion solvation data}.
\newblock \emph{\bibinfo{journal}{The Journal of Physical Chemistry A}}
  \textbf{\bibinfo{volume}{102}}, \bibinfo{pages}{7787--7794}
  (\bibinfo{year}{1998}).

\bibitem{flyvbjerg1989error}
\bibinfo{author}{Flyvbjerg, H.} \& \bibinfo{author}{Petersen, H.~G.}
\newblock \bibinfo{title}{Error estimates on averages of correlated data}.
\newblock \emph{\bibinfo{journal}{The Journal of Chemical Physics}}
  \textbf{\bibinfo{volume}{91}}, \bibinfo{pages}{461--466}
  (\bibinfo{year}{1989}).

\bibitem{behler2007generalized}
\bibinfo{author}{Behler, J.} \& \bibinfo{author}{Parrinello, M.}
\newblock \bibinfo{title}{Generalized neural-network representation of
  high-dimensional potential-energy surfaces}.
\newblock \emph{\bibinfo{journal}{Physical review letters}}
  \textbf{\bibinfo{volume}{98}}, \bibinfo{pages}{146401}
  (\bibinfo{year}{2007}).

\bibitem{wang2018deepmd}
\bibinfo{author}{Wang, H.}, \bibinfo{author}{Zhang, L.}, \bibinfo{author}{Han,
  J.} \& \bibinfo{author}{Weinan, E.}
\newblock \bibinfo{title}{Deepmd-kit: A deep learning package for many-body
  potential energy representation and molecular dynamics}.
\newblock \emph{\bibinfo{journal}{Computer Physics Communications}}
  \textbf{\bibinfo{volume}{228}}, \bibinfo{pages}{178--184}
  (\bibinfo{year}{2018}).

\bibitem{Wang2026Machine}
\bibinfo{author}{Wang, R.}, \bibinfo{author}{Meraz, V.~J.} \&
  \bibinfo{author}{Tiwary, P.}
\newblock \bibinfo{title}{Machine {Learning} {Driven} {Advances} in {Molecular}
  {Dynamics} of {Bulk} and {Interfacial} {Aqueous} {Systems}}.
\newblock \emph{\bibinfo{journal}{Chemical Reviews}}  (\bibinfo{year}{2026}).

\bibitem{schran2020committee}
\bibinfo{author}{Schran, C.}, \bibinfo{author}{Brezina, K.} \&
  \bibinfo{author}{Marsalek, O.}
\newblock \bibinfo{title}{Committee neural network potentials control
  generalization errors and enable active learning}.
\newblock \emph{\bibinfo{journal}{The Journal of Chemical Physics}}
  \textbf{\bibinfo{volume}{153}}, \bibinfo{pages}{104105}
  (\bibinfo{year}{2020}).

\bibitem{Kapil2022first}
\bibinfo{author}{Kapil, V.} \emph{et~al.}
\newblock \bibinfo{title}{The first-principles phase diagram of monolayer
  nanoconfined water}.
\newblock \emph{\bibinfo{journal}{Nature}} \textbf{\bibinfo{volume}{609}},
  \bibinfo{pages}{512--516} (\bibinfo{year}{2022}).

\bibitem{hunenberger2011single}
\bibinfo{author}{H{\"u}nenberger, P.} \& \bibinfo{author}{Reif, M.}
\newblock \emph{\bibinfo{title}{Single-ion solvation: experimental and
  theoretical approaches to elusive thermodynamic quantities}},
  vol.~\bibinfo{volume}{3} (\bibinfo{publisher}{Royal Society of Chemistry},
  \bibinfo{year}{2011}).

\bibitem{kathmann2011understanding}
\bibinfo{author}{Kathmann, S.~M.}, \bibinfo{author}{Kuo, I.-F.~W.},
  \bibinfo{author}{Mundy, C.~J.} \& \bibinfo{author}{Schenter, G.~K.}
\newblock \bibinfo{title}{Understanding the surface potential of water}.
\newblock \emph{\bibinfo{journal}{The Journal of Physical Chemistry B}}
  \textbf{\bibinfo{volume}{115}}, \bibinfo{pages}{4369--4377}
  (\bibinfo{year}{2011}).

\bibitem{smith2005free}
\bibinfo{author}{Smith, E.}, \bibinfo{author}{Bryk, T.} \&
  \bibinfo{author}{Haymet, A.~D.}
\newblock \bibinfo{title}{Free energy of solvation of simple ions:
  Molecular-dynamics study of solvation of cl- and na+ in the ice/water
  interface}.
\newblock \emph{\bibinfo{journal}{The Journal of chemical physics}}
  \textbf{\bibinfo{volume}{123}}, \bibinfo{pages}{034706}
  (\bibinfo{year}{2005}).

\bibitem{kelly2006aqueous}
\bibinfo{author}{Kelly, C.~P.}, \bibinfo{author}{Cramer, C.~J.} \&
  \bibinfo{author}{Truhlar, D.~G.}
\newblock \bibinfo{title}{Aqueous solvation free energies of ions and ion-
  water clusters based on an accurate value for the absolute aqueous solvation
  free energy of the proton}.
\newblock \emph{\bibinfo{journal}{The Journal of Physical Chemistry B}}
  \textbf{\bibinfo{volume}{110}}, \bibinfo{pages}{16066--16081}
  (\bibinfo{year}{2006}).

\bibitem{tomanik2020solvation}
\bibinfo{author}{Toman{\'\i}k, L.}, \bibinfo{author}{Muchov{\'a}, E.} \&
  \bibinfo{author}{Slav{\'\i}{\v{c}}ek, P.}
\newblock \bibinfo{title}{Solvation energies of ions with ensemble
  cluster-continuum approach}.
\newblock \emph{\bibinfo{journal}{Physical Chemistry Chemical Physics}}
  \textbf{\bibinfo{volume}{22}}, \bibinfo{pages}{22357--22368}
  (\bibinfo{year}{2020}).

\bibitem{duignan2017real}
\bibinfo{author}{Duignan, T.~T.}, \bibinfo{author}{Baer, M.~D.},
  \bibinfo{author}{Schenter, G.~K.} \& \bibinfo{author}{Mundy, C.~J.}
\newblock \bibinfo{title}{Real single ion solvation free energies with quantum
  mechanical simulation}.
\newblock \emph{\bibinfo{journal}{Chemical science}}
  \textbf{\bibinfo{volume}{8}}, \bibinfo{pages}{6131--6140}
  (\bibinfo{year}{2017}).

\bibitem{duignan2017electrostatic}
\bibinfo{author}{Duignan, T.~T.}, \bibinfo{author}{Baer, M.~D.},
  \bibinfo{author}{Schenter, G.~K.} \& \bibinfo{author}{Mundy, C.~J.}
\newblock \bibinfo{title}{Electrostatic solvation free energies of charged hard
  spheres using molecular dynamics with density functional theory
  interactions}.
\newblock \emph{\bibinfo{journal}{The Journal of Chemical Physics}}
  \textbf{\bibinfo{volume}{147}}, \bibinfo{pages}{161716}
  (\bibinfo{year}{2017}).

\bibitem{Freysoldt2009fully}
\bibinfo{author}{Freysoldt, C.}, \bibinfo{author}{Neugebauer, J.} \&
  \bibinfo{author}{Van~de Walle, C.~G.}
\newblock \bibinfo{title}{Fully ab initio finite-size corrections for
  charged-defect supercell calculations}.
\newblock \emph{\bibinfo{journal}{Physics Review Letters}}
  \textbf{\bibinfo{volume}{102}}, \bibinfo{pages}{016402}
  (\bibinfo{year}{2009}).

\bibitem{harscher1998inelastic}
\bibinfo{author}{Harscher, A.} \& \bibinfo{author}{Lichte, H.}
\newblock \bibinfo{title}{Inelastic mean free path and mean inner potential of
  carbon foil and vitrified ice measured with electron holography}.
\newblock \emph{\bibinfo{journal}{ICEM14, Cancun, Mexico}}
  \textbf{\bibinfo{volume}{31}}, \bibinfo{pages}{553--554}
  (\bibinfo{year}{1998}).

\bibitem{sellner2014matter}
\bibinfo{author}{Sellner, B.} \& \bibinfo{author}{Kathmann, S.~M.}
\newblock \bibinfo{title}{A matter of quantum voltages}.
\newblock \emph{\bibinfo{journal}{The Journal of Chemical Physics}}
  \textbf{\bibinfo{volume}{141}}, \bibinfo{pages}{18C534}
  (\bibinfo{year}{2014}).

\bibitem{yesibolati2020mean}
\bibinfo{author}{Yesibolati, M.~N.} \emph{et~al.}
\newblock \bibinfo{title}{Mean inner potential of liquid water}.
\newblock \emph{\bibinfo{journal}{Physical Review Letters}}
  \textbf{\bibinfo{volume}{124}}, \bibinfo{pages}{065502}
  (\bibinfo{year}{2020}).

\bibitem{Note1}
\bibinfo{note}{In the CP2K code, the electrostatic potential $V(x,y,z)$ is
  stored on a 3D uniform real-space grid, with columns along the $z$-direction
  distributed across different MPI processes. While $V(x,y,z)$ could in
  principle be written to a single file (e.g., a Gaussian cube file) for
  post-processing, this approach would require prohibitively large disk space.
  In our implementation, each process computes partial sums for its local
  $z$-columns, which are then sent to the master process for final summation
  and analysis.}

\bibitem{leung2010surface}
\bibinfo{author}{Leung, K.}
\newblock \bibinfo{title}{Surface potential at the air- water interface
  computed using density functional theory}.
\newblock \emph{\bibinfo{journal}{The Journal of Physical Chemistry Letters}}
  \textbf{\bibinfo{volume}{1}}, \bibinfo{pages}{496--499}
  (\bibinfo{year}{2010}).

\bibitem{cheng2009redox}
\bibinfo{author}{Cheng, J.}, \bibinfo{author}{Sulpizi, M.} \&
  \bibinfo{author}{Sprik, M.}
\newblock \bibinfo{title}{Redox potentials and p k a for benzoquinone from
  density functional theory based molecular dynamics}.
\newblock \emph{\bibinfo{journal}{The Journal of chemical physics}}
  \textbf{\bibinfo{volume}{131}}, \bibinfo{pages}{154504}
  (\bibinfo{year}{2009}).

\bibitem{prozorov2017off}
\bibinfo{author}{Prozorov, T.}, \bibinfo{author}{Almeida, T.~P.},
  \bibinfo{author}{Kov{\'a}cs, A.} \& \bibinfo{author}{Dunin-Borkowski, R.~E.}
\newblock \bibinfo{title}{Off-axis electron holography of bacterial cells and
  magnetic nanoparticles in liquid}.
\newblock \emph{\bibinfo{journal}{Journal of the Royal Society Interface}}
  \textbf{\bibinfo{volume}{14}}, \bibinfo{pages}{20170464}
  (\bibinfo{year}{2017}).

\bibitem{kusalik1994spatial}
\bibinfo{author}{Kusalik, P.~G.} \& \bibinfo{author}{Svishchev, I.~M.}
\newblock \bibinfo{title}{The spatial structure in liquid water}.
\newblock \emph{\bibinfo{journal}{Science}} \textbf{\bibinfo{volume}{265}},
  \bibinfo{pages}{1219--1221} (\bibinfo{year}{1994}).

\bibitem{ryckaert1977numerical}
\bibinfo{author}{Ryckaert, J.-P.}, \bibinfo{author}{Ciccotti, G.} \&
  \bibinfo{author}{Berendsen, H.~J.}
\newblock \bibinfo{title}{Numerical integration of the cartesian equations of
  motion of a system with constraints: molecular dynamics of n-alkanes}.
\newblock \emph{\bibinfo{journal}{Journal of computational physics}}
  \textbf{\bibinfo{volume}{23}}, \bibinfo{pages}{327--341}
  (\bibinfo{year}{1977}).

\bibitem{evans1985nose}
\bibinfo{author}{Evans, D.~J.} \& \bibinfo{author}{Holian, B.~L.}
\newblock \bibinfo{title}{The nose--hoover thermostat}.
\newblock \emph{\bibinfo{journal}{The Journal of chemical physics}}
  \textbf{\bibinfo{volume}{83}}, \bibinfo{pages}{4069--4074}
  (\bibinfo{year}{1985}).

\bibitem{Hoover1985Canonical}
\bibinfo{author}{Hoover, W.~G.}
\newblock \bibinfo{title}{Canonical dynamics: {Equilibrium} phase-space
  distributions}.
\newblock \emph{\bibinfo{journal}{Physical Review A}}
  \textbf{\bibinfo{volume}{31}}, \bibinfo{pages}{1695--1697}
  (\bibinfo{year}{1985}).

\bibitem{tuckerman2006liouville}
\bibinfo{author}{Tuckerman, M.~E.}, \bibinfo{author}{Alejandre, J.},
  \bibinfo{author}{L{\'o}pez-Rend{\'o}n, R.}, \bibinfo{author}{Jochim, A.~L.}
  \& \bibinfo{author}{Martyna, G.~J.}
\newblock \bibinfo{title}{A liouville-operator derived measure-preserving
  integrator for molecular dynamics simulations in the isothermal--isobaric
  ensemble}.
\newblock \emph{\bibinfo{journal}{Journal of Physics A: Mathematical and
  General}} \textbf{\bibinfo{volume}{39}}, \bibinfo{pages}{5629}
  (\bibinfo{year}{2006}).

\bibitem{kuhne2020cp2k}
\bibinfo{author}{K{\"u}hne, T.~D.} \emph{et~al.}
\newblock \bibinfo{title}{{CP2K: An electronic structure and molecular dynamics
  software package-Quickstep: Efficient and accurate electronic structure
  calculations}}.
\newblock \emph{\bibinfo{journal}{The Journal of Chemical Physics}}
  \textbf{\bibinfo{volume}{152}}, \bibinfo{pages}{194103}
  (\bibinfo{year}{2020}).

\bibitem{vandevondele2005quickstep}
\bibinfo{author}{VandeVondele, J.} \emph{et~al.}
\newblock \bibinfo{title}{Quickstep: Fast and accurate density functional
  calculations using a mixed gaussian and plane waves approach}.
\newblock \emph{\bibinfo{journal}{Computer Physics Communications}}
  \textbf{\bibinfo{volume}{167}}, \bibinfo{pages}{103--128}
  (\bibinfo{year}{2005}).

\bibitem{goedecker1996separable}
\bibinfo{author}{Goedecker, S.}, \bibinfo{author}{Teter, M.} \&
  \bibinfo{author}{Hutter, J.}
\newblock \bibinfo{title}{Separable dual-space gaussian pseudopotentials}.
\newblock \emph{\bibinfo{journal}{Physical Review B}}
  \textbf{\bibinfo{volume}{54}}, \bibinfo{pages}{1703} (\bibinfo{year}{1996}).

\bibitem{hartwigsen1998relativistic}
\bibinfo{author}{Hartwigsen, C.}, \bibinfo{author}{G{\oe}decker, S.} \&
  \bibinfo{author}{Hutter, J.}
\newblock \bibinfo{title}{Relativistic separable dual-space gaussian
  pseudopotentials from h to rn}.
\newblock \emph{\bibinfo{journal}{Physical Review B}}
  \textbf{\bibinfo{volume}{58}}, \bibinfo{pages}{3641} (\bibinfo{year}{1998}).

\bibitem{grimme2010consistent}
\bibinfo{author}{Grimme, S.}, \bibinfo{author}{Antony, J.},
  \bibinfo{author}{Ehrlich, S.} \& \bibinfo{author}{Krieg, H.}
\newblock \bibinfo{title}{A consistent and accurate ab initio parametrization
  of density functional dispersion correction (dft-d) for the 94 elements
  h-pu}.
\newblock \emph{\bibinfo{journal}{The Journal of chemical physics}}
  \textbf{\bibinfo{volume}{132}}, \bibinfo{pages}{154104}
  (\bibinfo{year}{2010}).

\bibitem{bussi2007canonical}
\bibinfo{author}{Bussi, G.}, \bibinfo{author}{Donadio, D.} \&
  \bibinfo{author}{Parrinello, M.}
\newblock \bibinfo{title}{Canonical sampling through velocity rescaling}.
\newblock \emph{\bibinfo{journal}{The Journal of chemical physics}}
  \textbf{\bibinfo{volume}{126}}, \bibinfo{pages}{014101}
  (\bibinfo{year}{2007}).

\bibitem{martyna1994constant}
\bibinfo{author}{Martyna, G.~J.}, \bibinfo{author}{Tobias, D.~J.} \&
  \bibinfo{author}{Klein, M.~L.}
\newblock \bibinfo{title}{Constant pressure molecular dynamics algorithms}.
\newblock \emph{\bibinfo{journal}{The Journal of chemical physics}}
  \textbf{\bibinfo{volume}{101}}, \bibinfo{pages}{4177--4189}
  (\bibinfo{year}{1994}).

\bibitem{guidon2010auxiliary}
\bibinfo{author}{Guidon, M.}, \bibinfo{author}{Hutter, J.} \&
  \bibinfo{author}{VandeVondele, J.}
\newblock \bibinfo{title}{Auxiliary density matrix methods for hartree- fock
  exchange calculations}.
\newblock \emph{\bibinfo{journal}{Journal of chemical theory and computation}}
  \textbf{\bibinfo{volume}{6}}, \bibinfo{pages}{2348--2364}
  (\bibinfo{year}{2010}).

\end{thebibliography}
\newpage

\begin{figure}
\centering
\includegraphics[width=0.5\linewidth]{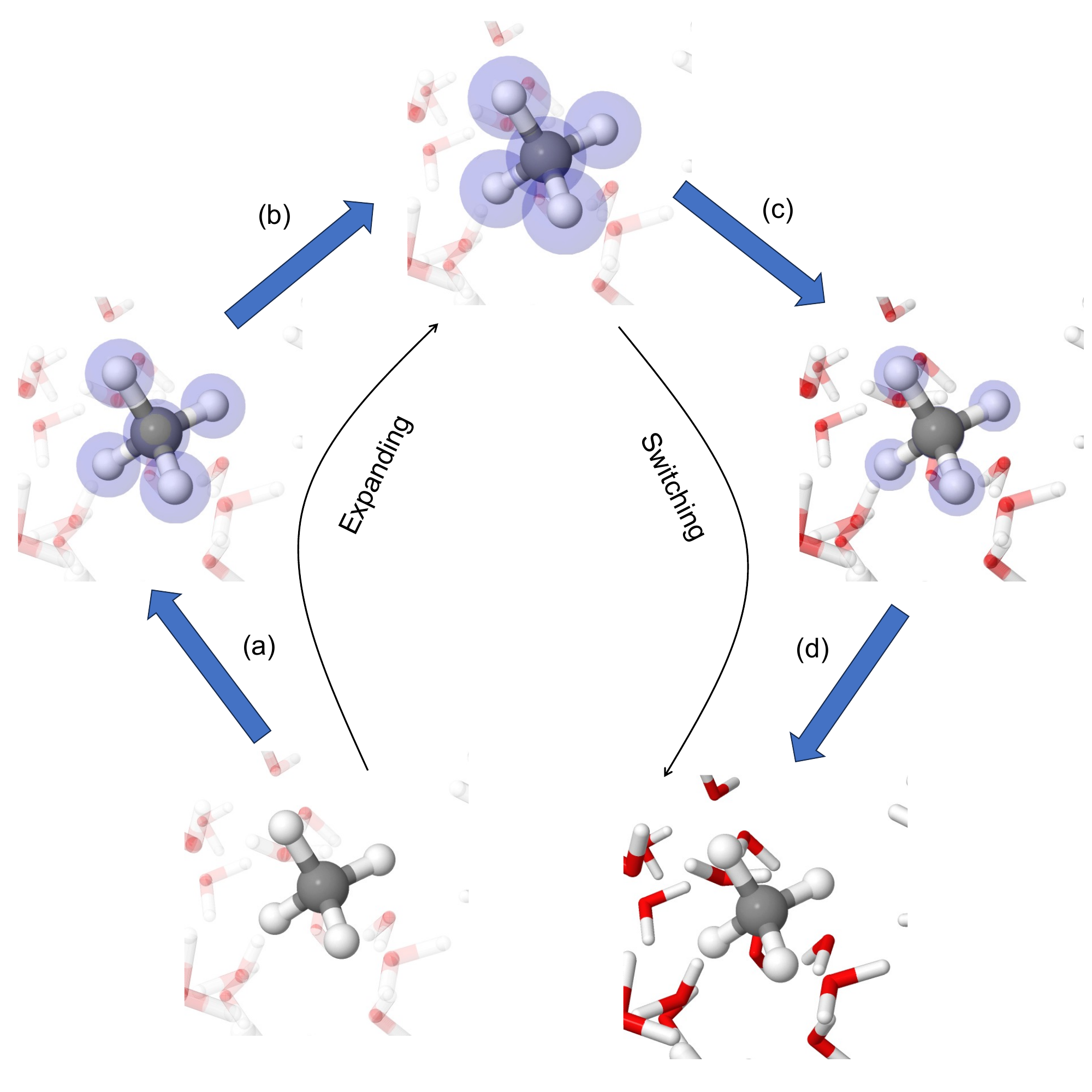}
\caption{Schematic illustration of the bubble method scheme. 
(a) In the absence of solute-solvent interactions, a spherical bubble forms and expands around each atom in the solute molecule. (b) The bubbles reach their predefined maximum radius. (c) The bubbles begin to shrink, and the solute-solvent interactions are gradually switched on. (d) The bubbles have fully collapsed, and the system has returned to its fully interacting state.}
\label{fig:demo}
\end{figure}

\begin{figure}
\centering
\includegraphics[width=.8\linewidth]{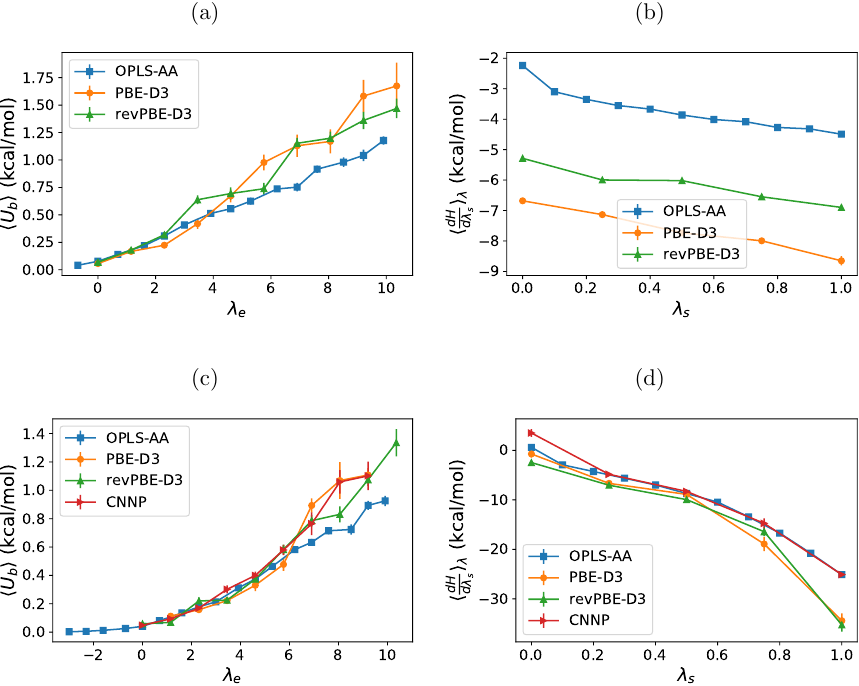}
\caption{Integrands used in the thermodynamic integration to calculate the solution free energy in the bubble method.
(a) Expanding process of CH$_4$(aq), (b) Switching process of CH$_4$(aq), (c) Expanding process of H$_2$O(aq), (b) Switching process of H$_2$O(aq). }
\label{fig:neutral}
\end{figure}

\begin{figure}
\centering
\includegraphics[width=.8\linewidth]{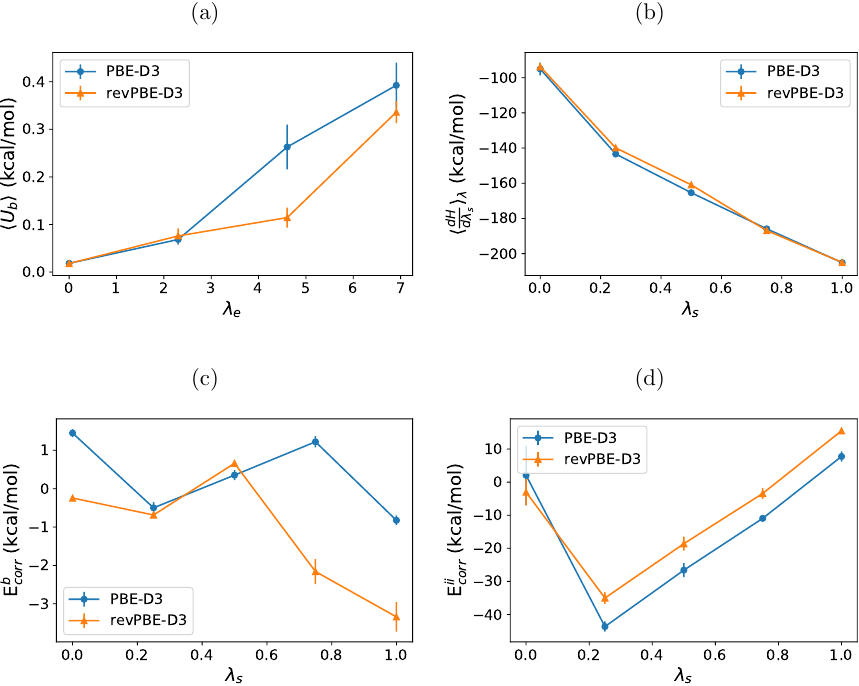}
\caption{Integrands used in the thermodynamic integration and corrections to calculate the intrinsic solution free energy of Na$^+$(aq) in the bubble method:
(a) expanding process, (b) switching process, (c) background correction, (d) ion-ion correction.}
\label{fig:Na_intrinsic}
\end{figure}

\end{document}